
\def \pb {\bar \pi}
\def \Hb {\bar H}
\def \half {{\textstyle {1 \over 2}}}
\def \PR {{\it Phys.~Rev.}}
\def \PRL {{\it Phys.~Rev.~Lett.}}
\magnification=1200
\nopagenumbers
\null \vskip -2truecm
\hbox to 6.5truein{\hfil NTUA 34/92}
\line {\hfil hepth@xxx/9210109}
\line {\hfil October 1992}
\vskip 2truecm
\centerline {\bf Lattice Integrable Systems of Haldane-Shastry Type}
\vskip 1.5truecm
\centerline {\bf Alexios P$.$ Polychronakos}
\vskip 0.8truecm
\centerline {Physics Department, National Technical University}
\centerline {GR-157 73 Zografou, Greece}
\vskip 0.5truecm
\centerline {and}
\vskip 0.5truecm
\centerline {Theoretical Physics Division, CERN}
\centerline {CH-1211 Geneva 23, Switzerland\footnote{$^*$}
{Address after December 1, 1992.}}

\vskip 1.5truecm
\centerline {\bf ABSTRACT}
\vskip 0.8truecm
\noindent
We present a new lattice integrable system in one dimension of the
Haldane-Shastry type. It consists of spins positioned at the
static equilibrium positions of particles in a corresponding classical
Calogero system and interacting through an exchange term with
strength inversely proportional to the square of their distance.
We achieve this by viewing the Haldane-Shastry system as a
high-interaction limit of the Sutherland system of
particles with internal degrees of freedom and identifying the same
limit in a corresponding Calogero system. The commuting integrals of
motion of this system are found using the exchange operator formalism.

\vskip 2truecm
\vfil
\eject
\baselineskip 0.8truecm

\footline={\hss\tenrm\folio\hss}
\pageno=2
Recently, the interest in spin systems of the Haldane-Shastry type as
well as in integrable systems of particles with internal degrees of
freedom has been revived [1-8]. The Haldane-Shastry model for spin chains
and its $SU(n)$
generalization consists of spins or, in general, $SU(n)$ color degrees
of freedom equally spaced around the unit circle with the hamiltonian [1]
$$
H = \sum_{i<j} {1 \over \sin^2 ({x_i - x_j \over 2})} P_{ij}
\eqno(1)$$
where $x_i$ are the positions of the spins and $P_{ij}$ is the
operator which exchanges the spins or colors of sites $i$ and $j$.
Haldane and Shastry found the antiferromagnetic ground state
wavefunction of the system, which is similar in form to the ground
state wavefunction of the Sutherland system of particles on the
circle [9], as well as all energy levels for the system.

Although the above
system was suspected to be integrable, and particular commuting
integrals of the motion were sporadically found [2,10], a complete proof
was lacking. Recently, however, Fowler and Minahan [11] showed the
integrability of the system and derived the conserved quantities
using a recently developed exchange operator formalism [12]. Their
approach consists of working initially with a system of $N$ bosons
with internal degrees of freedom and no kinematics which sit on the $N$
lattice sites and only allowing states with exactly one particle per
site. Then every operator which is invariant under particle
permutation must involve degrees of freedom on all lattice sites and
can thus be substituted with a corresponding lattice operator.
Integrability of this particle system, then, translates into
integrability for the lattice system. These authors, then, consider
the operators
$$
\pi_i = \sum_{j \neq i} { z_j \over z_{ij}} M_{ij}
\eqno(2)$$
where the indices $i,j$ now refer to particles, $z_i = \exp (2\pi i
x_i)$ and $z_{ij} = z_i - z_j$. $M_{ij}$ are the operators which
exchange the {\it positions} of particles, satisfying
$$
M_{ij} x_i = x_j M_{ij} ~,~~ M_{ij} x_k = x_k M_{ij}
{}~~~~{\rm (for}~i \neq k \neq j)
\eqno(3)$$
as well as the standard permutation group commutation relations
among themselves. The hamiltonian of the system is taken to be
$$
H = \sum_{i<j} {1 \over \sin^2 ({x_i - x_j \over 2})} M_{ij}
\eqno(4)$$
Using the commutation properties of $M_{ij}$ and $z_i$ one can
show that the quantities
$$
I_n = \sum_i \pi_i^n
\eqno(5)$$
commute among themselves and, if the lattice sites are equidistant,
they also commute with the hamiltonian. Therefore this system is
integrable. Each operator $M_{ij}$, now, acting on a bosonic state
translates into a spin exchange operator $\sigma_{ij}$ for the
particles [8]. Since the $I_n$ are symmetric under particle permutation,
every {\it particle} spin exchange operator they contain will
translate into a {\it site} spin exchange operator $P_{ij}$
and will reduce to the commuting conserved quantities
of the corresponding Haldane-Shastry lattice system.

The above operators $\pi_i$ considered by Fowler and Minahan are,
in fact, identical in form to the corresponding operators considered by this
author in the exchange operator formalism of the Sutherland problem [12], only
lacking an explicit kinetic term. These operators are
$$
\pb_i = p_i + il \sum_{j \neq i} \cot \Bigl({x_i - x_j \over 2}\Bigr) M_{ij}
+ l \sum_{j \neq i} M_{ij} = p_i -2l \pi_i
\eqno(6)$$
(from now on we use barred symbols to represent quantities for the
fully dynamical system of particles) while the hamiltonian is
$$
\Hb = \sum_i \half p_i^2 + \sum_{i<j} {l(l- M_{ij}) \over \sin^2 (x_i - x_j )}
\eqno(7)$$
If we rescale $\pb_i$ by a factor of $- \half l^{-1}$ and take the limit $l
\to \infty$ we see that the kinetic term drops and we recover $\pi_i$.
On the other hand, the leading term in $H$ in this limit (of order
$l^2$) becomes a nondynamical constant and can be subtracted away. The
highest nontrivial term then becomes exactly of the form (4).

We see therefore that the lattice system can be viewed as the
high-interaction limit of a corresponding Sutherland model. (The
strength of the two-body interaction is of order $l^2$.) We must be
careful, though. In fact, the limit $l \to \infty$ is the same as the
{\it classical limit} $\hbar \to 0$ (this can be seen by restoring
$\hbar$ into the problem) and, of course, the momentum does not become
irrelevant in the classical limit. The point is that $p_i$ as an
operator has an unbound spectrum and therefore cannot be neglected
no matter how large $l$ becomes. In order to consistently drop it we
must restrict our attention to a portion of the spectrum of $p_i$
of finite size. Since the dependence of the energy on the quanta of
momentum is of the form [9] $\half (n+l)^2$, where $n$ is the momentum
quantum number, already states with $n=1$ are an energy of order
$l$ higher than those for $n=0$ and thus we must consider states with
no momentum excitations. The internal degrees of freedom then remain
the only dynamical variables of the problem. This also means that in
the classical limit $l \to \infty$ the classical value of the
momentum is zero (since a classical excitation requires a large number
of quanta). Thus, the particles must lie at the positions of their
{\it static} classical equilibrium, which, for the Sutherland model, are
evenly spaced on the circle. This gives a natural explanation to the
fact that the system is integrable only when the lattice points are
taken to be equidistant [11].

The above suggests a natural generalization: to every integrable
system of particles with internal degrees of freedom corresponds an
integrable lattice system through an
appropriate ``classical" limit. In particular, there should be a
lattice system with inverse square interactions and the lattice
points positioned at the equilibrium positions of the Calogero system.
In the remaining of this paper we rigorously establish this fact.

Consider the $N$-body Calogero system with potential [13]
$$
V = \sum_i \half l^2 \omega^2 x_i^2 + \sum_{i<j} {l^2 \over x_{ij}^2}
\eqno(8)$$
In order for this system to have a nontrivial classical equilibrium
configuration at the $l \to \infty$ limit we took the strength of the
harmonic oscillator potential to scale as $l^2$,
else the particles will either collapse to the origin or fly away to
infinity. Then, following Fowler and Minahan, we consider a system of
particles with internal degrees of freedom with the hamiltonian
$$
H = \sum_{i<j} {1 \over x_{ij}^2} M_{ij}
\eqno(9)$$
where the particle positions $x_i$ are taken to minimize the above
potential. The parameter $\omega$ can be absorbed into a rescaling of
the particle positions $x_i$ which results in a mere rescaling of the
hamiltonian (9). We put it therefore equal to one, and the $x_i$ satisfy
$$
x_i - \sum_{j \neq i} {2 \over x_{ij}^3} = 0
\eqno(10)$$
Consider then the operators
$$
\pi_i = \sum_{j \neq i} {i \over x_{ij}} M_{ij}
\eqno(11)$$
These can be thought as the large-$l$ limit of the corresponding
operators $\pb_i$ defined in [12]
$$
\pb_i = p_i + l \pi_i
\eqno(12)$$
Similarly, the hamiltonian (9) can be thought as the large-$l$
limit of the full hamiltonian
$$
{\bar H} = {\bar H}_o + \sum_i \half l^2 x_i^2
\eqno(13)$$
$$
{\bar H}_o = \sum_i \half \pb_i^2 = \sum_i \half p_i^2 -l H + \sum_{i<j}
{l^2 \over x_{ij}^2}
\eqno(14)$$
after dropping trivial nondynamical terms of order $l^2$. Since the
operators $\pb$ commute [12], we immediately obtain
$$
[ \pi_i , \pi_j ] = 0
\eqno(15)$$
The commutation properties of $\pi_i$ with $H$ can be calculated
directly. All the labor can be saved, however, by taking the relation
$$
[ \pb_i , {\bar H}_o ] = 0
\eqno(16)$$
and expanding in powers of $l$. Since it holds for all $l$, each
term must separately vanish. The term of order $l^2$ gives
$$
[ p_i , \sum_{i<j} {1 \over x_{ij}^2} ] - [ \pi_i , H ] = 0
\eqno(17)$$
and we immediately obtain
$$
[ \pi_i , H ] = \sum_{j \neq i} {2i \over x_{ij}^3}
\eqno(18)$$
Again, following [12], consider the operators
$$
a_i^\dagger = \pi_i + i x_i \,\,,\,\,\,\, a_i = \pi_i - i x_i
\eqno(19)$$
$$
h_i = a_i^\dagger a_i = \sum_i ( \pi_i^2 + x_i^2 ) -
\sum_{j \neq i} M_{ij}
\eqno(20)$$
The commutation relations of $h_i$ can be directly deduced from the
corresponding relations for ${\bar h}_i$ to be
$$
[ h_i , h_j ] = -2 \, ( h_i M_{ij} - M_{ij} h_i )
\eqno(21)$$
and this means that the permutation symmetric quantities
$$
I_n = \sum_i h_i^n
\eqno(22)$$
commute among themselves. The proof is by now standard [12] and will
not be repeated here.

It remains to show that the $I_n$ commute with $H$. To this end we
have
$$
\eqalign{
[ \pi_i^2 , H ] &= \pi_i [ \pi_i , H ] + [ \pi_i , H ] \pi_i \cr
&= \pi_i \sum_{j \neq i} {2i \over x_{ij}^3} + \sum_{j \neq i}
{2i \over x_{ij}^3} \pi_i \cr}
\eqno(23)$$
and
$$
\eqalign{
[ x_i^2 , H ] &= \sum_{j \neq i} { x_i^2 - x_k^2 \over x_{ij}^2}
M_{ij} = \sum_{j \neq i} {x_i + x_j \over x_{ij}} M_{ij} \cr
&= -i x_i \pi_i -i \pi_i x_i \cr}
\eqno(24)$$
Finally, $H$ being permutation symmetric we have
$$
[ M_{ij} , H ] = 0
\eqno(25)$$
Putting everything together we get
$$
[ h_i , H ] = -i \left( x_i - \sum_{j \neq i} {2 \over x_{ij}^3} \right)
\pi_i -i \pi_i
\left( x_i - \sum_{j \neq i} {2 \over x_{ij}^3} \right)
\eqno(26)$$
We see that the quantity appearing in the parenthesis is exactly the
equation for classical equilibrium (10). Therefore, the quantities
$h_i$, and consequently also $I_n$, will commute with the hamiltonian
if the $x_i$ are chosen to correspond to the positions of Calogero
particles at rest.

The remaining of the argument is as in [11]. The $I_n$ will contain
strings of operators $M_{ij}$ which, when acting on totally symmetric
(bosonic) states become strings of operators $\sigma_{ij}$ in the reverse
order and those, in their turn, can be substituted by $P_{ij}$ operators.
These so-reduced operators ${\tilde I}_n$ then will constitute the
commuting integrals of motion of a lattice system with hamiltonian
as in (9) but with the spin exchange operator $P_{ij}$ appearing
instead of the position exchange operator $M_{ij}$.

The integrals obtained above should contain the hamiltonian itself,
but, just as in the Haldane-Shastry model, they do so in a nontrivial
way. In fact, since each $h_i$ in our model contains two exchange operators,
as opposed to only one in $\pi_i$ for the previous model, the form of
$I_n$ is more complicated. The lowest integral $I_1$ is
$$
I_1 = \sum_i h_i = \sum_i x_i^2 + \sum_{i \neq j} {1 \over x_{ij}^2}
- \sum_{i \neq j} M_{ij}
\eqno(27)$$
Apart from a nondynamical term (which is twice the classical rest
energy of the Calogero system) it consists only of a trivial exchange
operator. The non-trivial conserved quantities are contained in higher
$I_n$.

How different is our system from the Haldane-Shastry system?
Conceivably, the different form of the interaction in (9) could
conspire with the different lattice spacings to give something
similar. It is easy to see, however, that this is not the case.
Even the smallest nontrivial systems, $N=3$, are distinct: the
Haldane-Shastry system consists of three equal strength exchange
interactions, while our model consists of three interactions with
strengths at the ratio 4:4:1. Further, our system exhibits no analog of
the (lattice) translation invariance of the Haldane-Shastry system.
The dynamical properties of this system remain an interesting issue.

\vfil
\eject

{\centerline {\it Acknowledgements}
\vskip 0.4truecm

I would like to thank the National Technical University of Athens for its
hospitality and technical support during the writing of this paper.
I would also like to express my gratitude to the Greek Army, and
especially to Lieutenant Colonel E. Siakavellas, for giving me the
opportunity to do this work while fulfilling my military service obligations.

\vskip 0.8truecm

{\centerline {\bf REFERENCES}}
\vskip 0.4truecm

\item{[1]}
F.~D.~M.~Haldane, \PRL~{\bf 60} (1988) 635;
B.~S.~Shastry, \PRL~{\bf 60} (1988) 639.

\item{[2]}
V.~I.~Inozemtsev, {\it Jour.~Stat.~Phys.}~{\bf 59} (1989) 1143.

\item{[3]}
F.~D.~M.~Haldane, \PRL~{\bf 66} (1991) 1529.

\item{[4]}
H.~Kiwata and Y.~Akutsu, {\it Jour.~Phys.~Soc.~Jap.}~{\bf 61} (1992)
1441.

\item{[5]}
N.~Kawakami, \PR~{\bf B46} (1992) 1005 and 3191.

\item{[6]}
B.~S.~Shastry, \PRL~{\bf 69} (1992) 164.

\item{[7]}
Z.~N.~C.~Ha and F.~M.~Haldane, Princeton Preprint 1992, to appear in
Phys.~Rev.~B.

\item{[8]}
J.~A.~Minahan and A.~P.~Polychronakos, Virginia and Columbia Preprint
UVA-HET-92-04 and CU-TP-566.

\item{[9]}
B.~Sutherland, \PR~{\bf A4} (1971) 2019 and {\bf A5} (1972) 1372;
\PRL~{\bf 34} (1975) 1083.

\item{[10]}
F.~D.~M.~Haldane, private communication with M.~Fowler.

\item{[11]}
M.~Fowler and J.~A.~Minahan, Virginia Preprint UVA-HET-92-07.

\item{[12]}
A.~P.~Polychronakos, \PRL~{\bf 69} (1992) 703.

\item{[13]}
F.~Calogero, {\it Jour.~Math.~Phys.}~{\bf 10} (1969) 2191 and 2197;
{\bf 12} (1971) 419.

\end